\documentclass[10pt,preprint]{aastex}

\usepackage{epsf}

\def \etal   {\hbox{\it et~al.\/}}

\begin{document}

%
%

\title{Free-Free Spectral Energy Distributions of Hierarchically Clumped
	HII Regions}

\author{Richard Ignace\footnote{Currently at Department of
Physics and Astronomy, East Tennessee State University; Email:
ignace@mail.etsu.edu}~~\& Ed\ Churchwell}

\affil{
        Department of Astronomy,
        University of Wisconsin,
        5534 Sterling Hall,
        475 N.\ Charter St.,
        Madison, WI  53706-1582 }


%
%

\begin{abstract}

In an effort to understand unusual power-law spectral slopes observed
in some hypercompact HII regions, we consider the radio continuum energy
distribution from an ensemble of spherical clumps.  An analytic expression
for the free-free emission from a single spherical clump is derived.
The radio continuum slope (with $F_\nu \propto \nu^\alpha$) is governed
by the population of clump optical depths $N(\tau)$, such that (a)
at frequencies where all clumps are thick, a continuum slope of $+2$
is found, (b) at frequencies where all clumps are optically thin, a
flattened slope of $-0.11$ is found, and (c) at intermediate frequencies,
a power-law segment of significant bandwidth with slopes between these
two limiting values can result.  For the ensemble distribution, we adopt
a power-law distribution $N(\tau) \propto \tau^{-\gamma}$, and find
that significant power-law segments in the SED with slopes from $+2$
to $-0.11$ result only for a relatively restricted range of $\gamma$
values of 1 to 2.  Further, a greater range of clump optical depths for
this distribution leads to a wider bandwidth over which the intermediate
power-law segment exists.  The model is applied to the source W49N-B2
with an observed slope of $\alpha\approx +0.9$, but that may be turning
over to become optically thin around 2 mm. An adequate fit is found in
which most clumps are optically thin and there is little ``shadowing''
of rearward clumps by foreground clumps (i.e., the geometrical covering
factor $C \ll 1$).  The primary insight gained from our study is that
in the Rayleigh-Jeans limit for the Planck function that applies for
the radio band, it is the distribution in optical depth of the clump
population that is solely responsible for setting the continuum shape,
with variations in the size and temperature of clumps serving to modulate
the level of free-free emission.

\end{abstract}

\section{INTRODUCTION}

During the past few years a new class of super-compact HII regions has
been recognized via high resolution radio observations.  This class of
objects is now referred to as hypercompact (HC) HII regions (Gaume 1994;
Kurtz 2000).   HC HII regions are typically about ten times smaller and
about a hundred times denser than ultracompact (UC) HII regions (Kurtz
and Franco 2000; Kurtz 2002) and have emission measures typically $\geq
10^8$ pc cm$^{-6}$.  HC HII regions also have rising spectral energy
distributions (SEDs) from short cm to mm wavelengths (Hofner \etal\
1996; Kurtz 2002; Sewilo \etal\ 2003 and references therein) and often
have unusually broad radio recombination lines ($\geq 40$ km s$^{-1}$)
(Gaume 1994; Gaume \etal\ 1995; Shepherd \etal\ 1995; De Pree, Mehringer,
\& Goss 1997; Sewilo \etal\ 2003).  Some HC HII regions (see Sewilo \etal\
2003), like some UC HII regions (see Koo \etal\ 1996; Garay \etal\
1993; and Kurtz \etal\ 1999) are surrounded by extended low-density
ionized halos.  The most likely explanation for the halos is that  HC
and UC HII regions are highly clumped, thereby producing porous nebulae
to UV photons.  An example of a HC HII region with a power-law SED is
G75.78+0.34-H$_2$O, whose spectrum from 6~cm to 7~mm is
shown in Figure~\ref{fig1}.

HC HII regions are found in the vicinity of massive star formation and are
often coincident with strong H$_2$O masers.  Some HC HII regions appear
to be driving bipolar molecular outflows (Hofner \etal\ 1996; Shepherd
\etal\ 1998).  Sewilo \etal\ (2003) have suggested that HC HII
regions may represent an evolutionary stage between hot molecular cores
and UC HII regions.  During this period, rapid accretion onto the central
protostar shuts down, and a circumstellar HII region first becomes large
enough to be detected.  A property of at least some HC HII regions that concerns
this paper is their power-law SEDs at radio wavelengths.  The SEDs of
HC HII regions have spectral indices $\alpha$ between $+0.3$ to $+1.6$
($S_{\nu}\propto\nu^{\alpha}$) with typical values of $\alpha\approx +1$
from short cm to mm wavelengths. All HC HII regions seem to have rising
power spectra in the range $\approx 3.6$~cm to $\leq 3$~mm, but the slopes
may differ from source to source.

These spectral indices are especially interesting because they cannot
easily be explained by thermal or nonthermal radio continuum emission
for constant density nebulae.  They are too shallow for optically thick
thermal emission, too steep for optically thin thermal emission, and
the wavelength interval is too broad to represent the transition from
optically thick to thin thermal emission.  The spectral indices are
inconsistent with optically thin synchrotron emission; also, radio
recombination lines require that HC HII regions are thermal.
In this paper, we explore the possibility that the power-law SEDs of HC
HII regions might possibly be the result of hierarchial clumping.

It has been known for some time that power-law density gradients with
distance from the ionizing star of HII regions produce radio power-law
SEDs (Olnon 1975; Panagia \& Felli 1975; Wright \& Barlow 1975).
Olnon (1975) showed that the relationship between the slope of the SED
($\alpha$) and the slope of the power-law density gradient ($\omega$
where $n_{e}\propto r^{-\omega}$) is $\alpha=(2\omega-3.1)/(\omega-0.5)$
for $\omega>1.5$. Thus, for $\alpha=0.6$, $\omega=2.0$; for $\alpha=+1$,
$\omega=2.6$; and for $\alpha=+1.5$, $\omega=4.7$.  The $\alpha=0.6$,
$\omega=2.0$ parameters correspond to the classical values for a
constant velocity wind (Wright \& Barlow 1975; Panagia \& Felli 1975).
Olnon (1975) points out that when the density gradient has a power-law
dependence on radius, the slope of the SED will be determined by the
value of $\omega$ at the radius where the optical depth is about unity.
That is, the size of the effective radiating surface depends on both
the density gradient and frequency.   Hartmann \& Cassinelli (1977)
showed that a radial outflow whose velocity is a power-law with radius of
index $\beta$ has a density power-law dependence on radius of $\beta-2$,
resulting in a SED power-law ($S_{\nu}\propto\nu^{2/3}$ for a constant
velocity wind).  The density power-law index $\omega$ increases so rapidly
with $\alpha$ that $\omega$ is improbably large for $\alpha\geq +1$
($\omega\geq 2.6$).  In light of the observed density structure of HII
regions discussed above, it seems unlikely that real HII regions have such
steep and well-behaved density structures with radius, especially in the
very early stages of evolution expected for HC HII regions.  We therefore
investigate an alternate possible explanation for the observed radio
power-law SEDs of HC HII regions, namely hierarchal clumping of nebular gas.

As used here, ``hierarchial clumping'' refers to a region filled with
clumps of ionized gas having a range of sizes, temperatures, and optical
depths defined by power-law distributions.  There need not be a medium
in which all the clumps are embedded although such a structure could
be accommodated in our analysis.  A hierarchial clump distribution is
not the same as a fractal distribution which posits clumps within clumps
within clumps.	In a fractal structure the emergent SED is complicated by
the fact that every clump is embedded in clumps of larger size, whereas
in a hierarchically clumped structure the main complication arises when
clumps begin to shadow other clumps, otherwise one does not have to be
concerned with radiation transfer through a myriad of larger clumps.

The fact that the interstellar medium (ISM) seems to be clumped on all
observed size scales in HII regions, planetary nebulae, and neutral
atomic and molecular clouds is a strong motivation for the study
presented here.  High resolution Hubble Space Telescope (HST) images of
the Orion nebula (O'Dell 2001 and references therein) have revealed 
an array of small-scale, ionized structures down to the resolution limit
of the HST.  The small scale structures (clumps, filaments, knots,
etc) are easiest to recognize in Orion because of its proximity to us,
but high resolution observations of other HII regions such as M16
also indicate that they are composed of a complex of many clumps of
varying sizes (Hester \etal\ 1996). It is unlikely that the clumpy
structures in Orion and M16 are unique; rather, they probably indicate
that such structure is inherent in all HII regions.  Small scale ionized
clumps are also seen in planetary nebulae (O'Dell \etal\ 2002, 2003).
High resolution VLBI observations of Galactic HI absorption toward
quasars (Faison \& Goss 2001; Faison \etal\ 1998) show that very small
clumps (on the order of a few AU) exist in neutral Galactic HI clouds.
Extensive CO observations (e.g., Falgarone \& Phillips 1996; Elmegreen
\& Falgarone 1996; Falgarone \etal\ 1998) have clearly demonstrated the
existence of small-scale structures in molecular clouds.

The origin of the clumpy structure in the various phases of the ISM
is controversial and may have different explanations in different
environments.  For example, turbulence has been suggested by several
authors as the origin of structure in molecular and HI clouds (Elmegreen
\& Falgarone 1996; Falgarone \etal\ 1998; Lazarian \& Pogosyan 2000).  However, at
least some of the structure in HII regions may be due to hydrodynamical
instabilities and/or to pre-existing structure in the natal cloud of
an emerging HC HII region.  The reason for clumpy structures in the ISM
around massive stars is beyond the scope of this paper.  Here, the
observationally established clumpiness of the ISM plus the presence of
extended halos around UC and HC HII regions motivates our analysis of the
radio free-free spectra of hierarchically clumpy HC HII regions.

In the following section, we derive an analytic expression for the
free-free emission from a single spherical clump, and employ the result
to consider the radio SEDs from an ensemble of clumps.  In \S 3, the
model is applied to the source W49N-B2. A brief discussion of the model
and its results appear in \S 4, and concluding are remarks given in \S 5.

\section{FREE-FREE EMISSION FROM SPHERICAL CLUMPS}

\subsection{Emission from a Single Clump}

Let us begin our discussion by examining the spectrum of a single
spherical clump of uniform density, temperature, and ionization.
A closely related problem has been solved by Osterbrock (1974).  Here the
result is expressed in a slightly different form in order to determine
the unresolved flux of emission from an ensemble of clumps.

We begin with a discussion of the total optical depth $\tau$ along any
ray that passes through a clump.  Its value is given by

\begin{equation}
\tau(p) = 2\kappa_\nu(\rho,T)\,\rho\,z(p),
\end{equation}

\noindent where $\kappa_\nu$ is the opacity, $\rho$ the clump density,
$T$ the clump temperature, $p$ the impact parameter of the ray, and
$z=\sqrt{R^2-p^2}$ the total length of the ray through the clump, with $R$
being the clump radius.  If we may further assume a constant source function
$S_\nu$ for the clump emission, then the emergent intensity $I_\nu$
from the clump along this ray is

\begin{equation}
I_\nu = S_\nu\,\left[1-e^{-\tau(p)}\right].
\end{equation}

\noindent Correspondingly, the total flux emitted by an unresolved clump
will be given by

\begin{equation}
F_\nu = \frac{2\pi}{D^2}\,S_\nu\,\int_0^R\,\left[1-e^{-\tau(p)}\right]\,
	p\,dp
\end{equation}

Through a change of variable, whereby the integral is evaluated in $z$
instead of $p$, the preceding expression can be solved to obtain

\begin{equation}
F_\nu = \frac{2\pi\,R^2}{D^2}\,S_\nu\,\left\{ 1 - \frac{2}{t^2}\,
\left[ 1 - \left(1+t\right)\,e^{-t} \right] \right\},
	\label{eq:clumpflux}
\end{equation}

\noindent where

\begin{equation}
t = 2\kappa_\nu(\rho,T)\,\rho\,R,
\end{equation}

\noindent is the optical depth along the clump's diameter.
Equation~(\ref{eq:clumpflux}) has the correct limits, in that for a very
thick clump with $t \gg 1$, the flux depends on the source
function and clump cross-section with $F_\nu \approx
\pi\,S_\nu\,R^2/D^2$, and for a thin clump with $t \ll 1$,
the flux expression reduces to $2\pi\,t\,S_\nu\,R^2/3D^2$.
Substituting in for $t$, one finds that optically thin clumps
have a flux given by the product of the clump volume and the emissivity,
as expected.

For the case at hand, we shall consider only free-free emission at long
wavelengths appropriate for the radio band.  The source function
will be Planckian, with $S_\nu = B_\nu = 2kT\nu^2/c^2$, and the product
of opacity and density will be given by

\begin{equation}
\kappa_\nu(\rho,T)\,\rho = 0.018\, Z_i^2\,T^{-3/2}\,g_\nu\,
	\nu^{-2}\,\frac{\rho^2}{\mu_i\,\mu_e\,m_H^2}\;\,{\rm cm^{-1}},
	\label{eq:kappa}
\end{equation}

\noindent where $Z_i$ is the rms ion charge, $g_\nu \approx
(\nu/1~{\rm GHz} )^{-0.11}$ is the Gaunt factor, $\mu_i$ and $\mu_e$
are the ion and electron mean molecular weights, $m_H$ is the mass of
hydrogen, $T$ is in K, $\nu$ is in Hz, and $\rho$ is in g~cm$^{-3}$.
Equation~(\ref{eq:kappa}) is based on an expression appearing in \S 5.9
of Allen's Astrophysical Quantities (4$^{th}$ ed., Cox 2000).

\subsection{Emission from a Distribution of Clumps}

In considering the free-free emission from an ensemble of clumps, it is
necessary first to make a few remarks about the emission from a single
clump as given in equation~(\ref{eq:clumpflux}) and expressions following.
First, the basic variables for the emission include $R$, $T$, $\rho$,
$t$, and $\nu$, but these are not independent.  In particular,
$t$ depends on all 4 of the other variables.  Moreover, density
is the only variable that does not appear explicitly in the flux equation.

We choose to ignore density in the discussion to follow and consider
the flux to be a function of $R$, $T$, $t$, and $\nu$.
Moreover, our discussion concerns free-free emission in the radio band,
and already, it has been noted that we shall assume the Rayleigh-Jeans
limit for $B_\nu$, the source function.  Folding these assumptions into
equation~(\ref{eq:clumpflux}), we have that

\begin{equation}
F_\nu = \frac{4\pi\,k\,T\,R^2}{c^2\,D^2}\,\nu^2\,G(t),
	\label{eq:radioflux}
\end{equation}

\noindent where

\begin{equation}
G(t) = \left\{ 1 - \frac{2}{t^2}\,\left[ 
	1 - \left(1+t\right)\,e^{-t} \right] \right\}.
\end{equation}

\noindent We thus come to a remarkable conclusion.  In the treatment at
hand, $\nu$ is a spectral variable.  A distribution of clumps will thus
depend on $R$, $T$, and $t$, but it is only $t$ that depends on frequency.
This means that if $R$, $T$, and $t$ can be treated as independent
variables for describing the distribution of emitting clumps, it is
only the distribution in $t$ that can affect the {\it spectral shape},
whereas the distributions in $R$ and $T$ can influence only the {\it
flux scale}.  Consequently, we shall not consider distributions in $R$
and $T$ and replace their appearance with average quantities.  For example
with a distribution of $N_{\rm cl}$ clumps at fixed diameter optical
depth $t$ (allowing for different temperatures, sizes, and densities),
equation~(\ref{eq:radioflux}) would become

\begin{equation}
F_\nu = \frac{4\pi\,\nu_0^2\,k\,\langle T \rangle \,\langle R^2 \rangle }
	{c^2\,D^2}\,N_{\rm cl}\,f^2\,G(t),
	\label{eq:radio2flux}
\end{equation}

\noindent where we have introduced a normalized frequency $f =\nu/\nu_0$
with $\nu_0$ a fiducial value.

In Figure~\ref{fig2}, a plot of $G$ versus $t$ is shown.  At small optical
depth, $G \propto t$, whereas at high optical depth, the function is given
by $G\approx 1$.  This figure shows how clumps of different optical depths
will contribute in a large ensemble.  However, noting that $t \propto
f^{-2} \,g_\nu \propto f^{-2.11}$, Figure~\ref{fig2} also suggests how a
single clump contributes to the ensemble total at different frequencies.
From this perspective low $t$ corresponds to high $f$, and vice versa.
The implication is that for any given clump, there will always be
frequencies at which the clump is thin and frequencies where it is thick.

Thus equation~(\ref{eq:radioflux}) is found to have two important limits.
For clumps that are very thick (i.e., $t \gg 1$), the resultant SED will
be proportional to $f^2\,G \propto f^2$.  If all the clumps are thin,
then the SED will approach $f^2\,G \propto f^{-0.11}$ which is the
scaling for the Gaunt factor in the radio band.  The variation between
thick and thin thus gives a power-law range between $-0.11$ and $+2$, so
that in principle, one can imagine a frequency ``window'' over which the
SED may approximate a power-law with a slope that is intermediate between
these two limiting values. Note that we are ignoring contributions from
the central star because the radio continuum emission of the star will
be dwarfed by the contribution from the HII region; however, the star
and its immediate circumstellar component may start to compete with the
extended emission at the shorter mm and far-IR wavelengths.

We now introduce a power-law number distribution in terms of the optical
depth along the clump diameter:

\begin{equation}
N(t) = N_0\,t^{-\gamma},
\end{equation}

\noindent such that the total number of clumps $N_{\rm cl}$ is given by

\begin{equation}
N_{\rm cl} = N_0\,\int_{t_1}^{t_2}\, 
	t^{-\gamma}\, dt,
	\label{eq:Ncl}
\end{equation}

\noindent and $N_0$ is a normalization constant.  Now mean physical
parameters such as $\langle T \rangle$ and $\langle R^2 \rangle$ are
defined with reference to this optical depth distribution.  Care must
be taken in handling this distribution, since $t = t(f)$.  One must
ensure that the total number of clumps $N_{\rm cl}$ is the same for all
frequencies, requiring that the normalization constant is a function of
frequency, which is given by

\begin{equation}
N_0 (f) = (1-\gamma)\,N_{\rm cl}\,\left[ \frac{1}
	{t_2^{\gamma-1}} - \frac{1}{t_1^{\gamma-1}}
	\right]^{-1} \propto g_\nu^{\gamma-1}\,f^{2(1-\gamma)}.
\end{equation}

\noindent Thus the normalization can be a fairly strong function
of frequency.  

The total flux of emission from the ensemble of unresolved clumps is
given by the following integral expression:

\begin{equation}
F_{\rm tot}(f) = N_{\rm cl}\, F_0\,f^2\,
	\frac{\int_{t_1}^{t_2} \,
	N(t)\,G(t)\,dt} 
	{\int_{t_1}^{t_2}\,N(t)\,dt},
	\label{eq:distflux}
\end{equation}

\noindent where

\begin{equation}
F_0 = \frac{4\pi\,\nu_0^2\,k\,\langle T \rangle \,\langle R^2 \rangle }
	{c^2\,D^2}.
	\label{eq:F0}
\end{equation}

\noindent Expression~(\ref{eq:distflux}) indicates that the total flux from
the distribution of clumps is equivalent to the average flux times
the total number of clumps in the volume.  Implicit is that there
is no shadowing of one clump by another.  Our approach does not
allow for emission by one clump to be absorbed by another clump and
re-processed.  

An expression of the shadowing effect can be estimated in terms of a
covering factor, $C$.  We define this to be the projected area of all
the clumps, divided by the projected area of the volume in which the
clumps reside.  Suppose that the volume is spherical of radius $L$,
then we have that

\begin{equation}
C = \frac{A_{\rm tot}}{\pi\,L^2} = N_{\rm cl}\,\frac{\langle R^2 \rangle}
	{L^2}.
	\label{eq:cover}
\end{equation}

\noindent Again, our analysis is only valid if the covering factor is
less than unity.  It is certainly clear that $\langle R^2\rangle /L^2
< 1$ must hold, otherwise one does not really have a distribution of
clumps, but a volume-filling cloud.  

As an illustration of how $C$
depends on the clump distribution, we assume a power law distribution
of sizes given by $N(R) \propto R^{-\beta}$, bounded by $R_{\rm max}$
and $R_{\rm min}$.  In Figure~\ref{fig3}, we plot the covering factor
as a function of $\beta$ for different values of $R_{\rm max}^2/L^2 <
1$ and ratios $R_{\rm min}/R_{\rm max}$ as indicated in each panel of
the figure.  The value of $N_{\rm cl}$ is varied to give the different
curves appearing in each panel, with the lowest curve always for $N_{\rm
cl}=1$, and the highest curve for $N_{\rm cl}=10^7$.
although our analysis does not explicitly determine $\beta$, $R_{\rm min}$,
or $R_{\rm max}$, the application of our method does allow inference of
$C$, that can constrain the parameters that describe the clump size
distribution.

\subsection{Limits for the Clump Distribution}

It is worthwhile to consider the limiting properties of clumps that
might influence the model SEDs.  One such example is that clumps that are
sufficiently massive can collapse under their own self-gravity.  This is
behavior that we seek to avoid, so we need to estimate its domain.  If we
take collapse to occur when the self-gravity of the clump exceeds its
thermal internal energy, and apply our assumptions of constant density
and temperature, then a limit to the density for a given clump radius
can be derived.  The governing expressions for collapse are

\begin{equation}
\frac{3}{2}\,N\,k\,T_0 - \frac{3\,G\,M^2}{5\,R} < 0,
\end{equation}

\noindent where the clump mass is $M=4\pi \rho R^3/3$, and the
number of particles is $N=M/m$, for an average particle mass of $m$.
Re-arranging the expression, one finds an upper limit to the clump
size for a given density:

\begin{equation}
R < \sqrt{ \frac{15\,k\,T_0}{4\pi\,G\,m\,\rho} } .
\end{equation}

\noindent For a reasonable cold cloud temperature of $T_0=30$~K, and for
a quite large cloud number density of $10^9$ cm$^{-3}$ corresponding to
a density of $\rho\approx 10^{-15}$ g~cm$^{-3}$, the cloud will begin
to collapse if its size exceeds a value of about 500~AU.  For a radio
measurement around 10~GHz, the implied clump optical depth is about
$10^6$.  However, this density is extreme, and values more typical of
dense clouds will be about 3 orders of magnitude smaller.  Since $R <
\rho^{-1/2}$, the upper limit to the scale of dense clouds mushrooms to
about 0.07~pc, with a corresponding optical depth at 10~GHz of nearly 100.
But a size of 0.07~pc is comparable to the size (if not larger) of an
entire HC~HII region.  Consequently, the condition of cloud
collapse does not provide a significant physical constraint on the
distribution of cloud sizes, nor the cloud optical depths.

However, there is another condition that provides a limit to the size
and density of the clump.  For a clump to contribute to the observed
spectrum, it must survive the strong UV photo-ionizing radiation field
of the central source, otherwise the clump will have evaporated to meld
with the lower density diffuse gas of the region.
We estimate this limit as follows.  Let $\rho$ be the total density of
the clump, including neutrals $\rho_{\rm n}$ and ions $\rho_{\rm i}$.
We imagine that the star ``turns on'', producing an ionization
front for every clump.  The differential equation governing the
relative growth of ions in the clump will involve a competition between
the ionizing radiation field of the central star, which we represent with
$\dot{N}_\gamma$ for the number of ionizing photons emitted each second,
and the recombination of the gas, which is represented by $\dot{N}_{\rm
rec}$ for the number of recombinations each second.

It is convenient to work with number densities $n$ and particle
numbers $N$ for the neutral and ionized gas.  The gas is assumed to
be pure hydrogen with $n_H=\rho/m_H$.  The particle number is related
to the number density and volume via $N=n\,V$, where $V$ is the volume
appropriate for the species being considered (i.e., if $n$ refers to the
neutral gas, then $V$ is the volume in the clump with neutral hydrogen).
We take the evolution of the clump ionization as

\begin{equation}
\dot{N}_{\rm i} = {\cal W}(R,r)\,\dot{N}_\gamma-\dot{N}_{\rm rec},
\end{equation}

\noindent where ${\cal W}(R,r) = R^2/4r^2$ is the geometrical dilution
factor (in this case, the solid angle of the clump as seen from the vantage
of the central star at distance $r$ from the clump), and hence is the
fraction of the ionizing radiation from the central source that impinges
on the clump.  We further identify

\begin{equation}
\dot{N}_{\rm rec} = n_{\rm i}\,n_{\rm e}\,\alpha_{\rm rec}\,V_{\rm ioniz},
\end{equation}

\noindent where $\alpha_{\rm rec}$ is a recombination coefficient in
cm$^{3}$ s$^{-1}$, $n_{\rm i}=n_{\rm e}$, and $V_{\rm ioniz}$ references
the volume of ionized gas.  It is assumed that the total volume of the
clump does not change.  For simplicity, ionized particles are not allowed
to leave the clump, nor do ions and neutrals mix, with the consequence
that $n_{\rm i} = n_H$ in the relevant volumes.

Incorporating these assumptions and definitions, the equation for the
clump ionization becomes

\begin{equation}
\dot{N}_{\rm i} = {\cal W}(R,r)\,\dot{N}_\gamma-\alpha_{\rm rec}\,
	n_H\,N_{\rm i}.
\end{equation}

\noindent This is a standard inhomogeneous first order linear differential
equation with the solution

\begin{equation}
N_{\rm i}(t) = \frac{{\cal W}\,\dot{N}_\gamma}{\alpha_{\rm rec}\,n_H}
	\,\left(1-e^{-\alpha_{\rm rec} n_H t} \right).
\end{equation}

\noindent As an order of magnitude estimate, $\alpha_{\rm rec}\sim 10^{-13}$
cm$^3$ s$^{-1}$ is typical, and $n_H  \sim 10^6$ cm$^{-3}$ might
represent a typical cool clump.  Thus the exponential term of the
solution vanishes after $t \gtrsim 10^7$ s, or a few months.  
Thus an equilibrium condition is quickly met.

Now, a clump is said to be photo-evaporated (i.e., ``destroyed'') when
the total volume has become ionized, with $V_{\rm ioniz} = V = 4\pi\,R^3/3$.
Setting $N_{\rm i} = n_H V$ and substituting for $\cal{W}$, one
obtains the following condition for the survival of a clump:

\begin{eqnarray}
n_H^2\, R & \ge & \frac{3\dot{N}_\gamma}{16\pi\,r^2\,\alpha_{\rm rec}\,} \\
 & \gtrsim & 3 \times 10^{27}\,\left(\frac{\dot{N}_\gamma}{10^{49}~{\rm s}^{-1}}
	\right)\,\left(\frac{1000~{\rm AU}}{r}\right)^2\,\left(
	\frac{10^{-13}~{\rm cm^3\,s^{-1}}}{\alpha_{\rm rec}} \right)\;{\rm 
	cm}^{-5}.
\end{eqnarray}

\noindent The nominal values of $\dot{N}_\gamma = 10^{49}$ s$^{-1}$
and $r=1000$~AU were motivated by the application to W49N--B2 to be
discussed in \S 3 (see also Tab.~\ref{tab1}).

Interestingly, this lower limit can be converted to a minimum free-free
optical depth that is required for clump survival.  Recall that the
clump optical depth along its diameter is $t = 2\kappa\rho R$.
Noting that $n_{\rm i}\,n_{\rm e} = n_H^2$, and evaluating the opacity
for a frequency of $\nu=10^{10}$~Hz, the minimum optical depth for clump
survival is

\begin{equation}
t \gtrsim  1,
\end{equation}

\noindent evaluated for a temperature of $T=10,000$~K.  This limit is
useful because it effectively truncates the $N(t)$ distribution,
which as shown in the following section, can have ramifications for the
bandwidth over which intermediate power-law segments will exist.

\subsection{Parameter Study}

In Figure~\ref{fig4} calculations of model SEDs are shown for
a range of $\gamma$ values as indicated.  Recall that $N\propto
t^{-\gamma}$ so that for $\gamma<1$, thicker clumps dominate
the population and for $\gamma>1$, thinner clumps dominate
the population.  In each panel the solid line is the normalized SED
with values corresponding to the ordinate at left.  Note that for
$f=1$, the lower and upper limits to the optical depth
distribution are 0.01 and 100, respectively.  The dashed line is the
local power-law slope (i.e., first derivative of the SED), with ordinate
to the right.  The dotted line is the variation of the power-law index
in normalized frequency (i.e., the second derivative of the SED).
Although no separate axis is given for the dotted curve, its value is
generally negative because the SEDs are concave down, and zero wherever
the SED is linear in the log-log plot of Figure~\ref{fig4}.

The turnover from a thick spectrum to one that is thin has a minimum
in the curve for the second derivative.  Generally, a single minimum
indicates a smooth and continuous turnover from a thick to a thin
spectrum.  However, the case of $\gamma=+1$ shows a departure from this
trend, displaying two downward peaks, indicating 
three distinct portions of the spectrum that exhibit a power-law.
In fact, the two peaks define a frequency ``window'' or band for an
intermediate power-law with a slope between $+2$ and $-0.11$.

Figure~\ref{fig5} shows more SED calculations for a finer grid of
$\gamma$ values, focussing on the range of $\gamma=1$ to 2.  In this
range intermediate power-law segments persist with windows of about 1~dex
in frequency.  At $\gamma=1$ the slope of the intermediate power-law is
high at about $\alpha\approx 1.6$, but with increasing $\gamma$, the value
of $\alpha$ drops.  The frequency windows are narrower for $\gamma=1$
and 2.  As seen in Figure~\ref{fig4}, the SED turnover is smooth
and regular for $\gamma$ values outside the range of 1--2.
Hence, we come to the interesting conclusion that a rather restricted
range of $\gamma$ values are suitable for producing intermediate
power-laws of significant bandwidth.

To illustrate better how intermediate power-law segments arise,
Figure~\ref{fig6} shows a model SED using a value of $\gamma=1.5$.
The solid line is the SED from the clump ensemble.  The dotted lines
are for contributions to the net spectrum from clumps of different
optical depths (artificially shifted upward for clarity of display).
The SEDs of individual clumps show a power-law segment of $f^2$
at low frequency and one of $f^{-0.11}$ at high frequency,
the two joined by a fairly rapid turnover.  The different combination of
clumps with different optical depths at the fiducial frequency $\nu_0$
serves to ``spread out'' the turnovers of the individual clump SEDs,
so as to produce in the net SED a significant power-law segment that has
a slope intermediate between $+2$ and $-0.11$.

It is crucial to consider how the lower and upper optical depth scales
for the clump distribution influence the SED.  Figure~\ref{fig7} shows
more models of the case $\gamma=1.5$, now with the upper limit $t_2$ fixed
at 100, and the lower limit $t_1$ varied from 0.0001 to 10 as indicated.
As the lower limit $t_1$ approaches the upper limit value, the bandwidth
of the intermediate power-law grows progressively narrower, nearly to
disappear in the final panel.  This is reasonable, because as $t_1$
approaches $t_2$, the distribution of clump optical depths becomes
progressively narrower, eventually to approach a $\delta$-function.
Figure~\ref{fig8} shows a similar kind of plot, but now $t_1$ is fixed
at a value of 0.001, and the upper limit $t_2$ is decreased by factors of
10 from 10,000 down to 0.1.  As $t_2$ approaches $t_1$, the intermediate
power-law segment again narrows.

Conclusions from varying the upper and lower optical depth limits
are as follows.  (a) When the ratio $t_2/t_1$
is near unity, the resulting SED is essentially that of a
single clump.  (b)  As the ratio is made large, an intermediate
power-law segment can arise for certain values of $\gamma$,
and large ratios of $t_2/t_1$ maximize the
frequency window over which this segment persists.  And (c) assuming
a significant power-law segment exists,
changing the individual values of $t_1$ and $t_2$
shifts the power-law segment in frequency.  Increasing $t_2$
shifts the window to higher frequency, whereas decreasing its value
results in a shift to lower frequencies.

Next, we use the experience gained from our parameter study to
model the spectrum of the HC~HII region W49N-B2 which exhibits
an observed radio SED characterized by a power-law index of $+0.9$.

\section{APPLICATION TO W49N--B2}

The thermal radio source W49A, first observed by Westerhout (1958),
lies at a distance of 11.4 kpc (Gwinn, Moran, \& Reid 1992) and is
composed of two main components W49A-N and W49A-S which are generally
referred to simply as W49N and W49S.  With high spatial resolution, these
two main components resolve into multiple, luminous radio components.
Dreher \etal\ (1984) and De Pree \etal\ (2000) resolve W49N into nine or
more very compact and luminous radio-continuum components.  Of the nine
sources identified by De~Pree \etal\ (2000) in W49N, seven have broad
radio recombination lines ($\ge 40$~km~s$^{-1}$) and rising spectra
in the wavelength range 13 mm to 3 mm.  We have selected one of the
brightest of these, W49N-B2, from De Pree \etal\ (2000) to attempt to
fit its SED from 13~mm to 3~mm (spectral index of $\alpha = +0.9$).
This compact, luminous radio source has a measured radius of 0.007 pc,
an rms electron density of $1.3\times 10^6$~cm$^{-3}$, an emission
measure of $2.4\times 10^{10}$~pc~cm$^{-6}$, and an ionizing photon flux
of $\ge 2\times 10^{49}$ s$^{-1}$, corresponding to an equivalent
ZAMS spectral type of O5.5 or hotter (De Pree \etal\ 2000).  W49N-B2 has
a shell-like morphology with lower brightness extensions to the NE and
SW (see Fig.~2 of De Pree \etal\ 2000).  A water maser has also been
detected $\sim 1''$ to the north of the W49N-B sources.

To model the spectrum of W49N-B2, we also include a 1.4~mm observation
from Wilner \etal\ (2001).  At this wavelength the calibration is much
more difficult.  We have used the radio image appearing in their Figure~1
(ii) to infer a lower limit of 400~mJy at 1.4~mm, that appears as a point
with an upwardly directed arrow in the lower panel of Figure~\ref{fig9}.
Although earlier work suggested a power-law slope of $\alpha\approx
0.9$ around 7~mm (appearing as the dotted line in the lower panel of
Fig.~\ref{fig9}), the lower limit of Wilner \etal, in conjunction with
their discussion of adjacent sources, suggestst that the spectrum is
becoming optically thin around 2~mm.

Although Figure~\ref{fig9} indicates that a satisfactory match to the
shape of the SED can be obtained with our model for a clumpy medium,
it remains for us to determine if we can match the scale of the
observed flux.  From equation~(\ref{eq:F0}) for the parameter $F_0$
and equation~(\ref{eq:cover}) for the definition of the covering factor,
one obtains

\begin{equation}
F_0 = \frac{4\pi\,k\,\langle T \rangle\,L^2}{\lambda_0^2\,D^2}\,C.
\end{equation}

\noindent We introduce a parameter $\Lambda$ as

\begin{equation}
\Lambda = f^2\,\frac{\int \, N(t)\,G(t)
	\,dt}{\int\, N(t)\,dt},
\end{equation}

\noindent so that the source flux can be re-expressed as

\begin{equation}
F_\nu = F_0\,\Lambda(f).
\end{equation}

\noindent Using values from Table~\ref{tab1} for W49N-B2, and
assuming that $\langle T \rangle=8000$ K, we have evaluated $F_0$ at
the wavelength $\lambda_0=7$ mm, which gives $F_0=1080\, {\rm mJy}\,
\times C$.  The observed flux level at 7~mm is $F_\nu = 550$ mJy.
From Figure~\ref{fig9}, we find a value of $\log \Lambda_0 = +0.58$
at $\log f=\log \nu/\nu_0 = 0.92$.  Combining these numbers yields
a covering factor of $C\approx 0.15$, which is quite small indicating
that shadowing of clumps by other clumps is not a major concern for this
source.  In other words, foreground clumps are not absorbing emission
from rearward clumps.

Indeed, the shadowing effects implied by $C\approx 15\%$ should be
considered an upper limit for the following reason.  Our covering
factor is defined purely in terms of geometry.  If the majority of
clumps are optically thin, then the ``effective'' covering factor may
actually be less, since shadowing by thin clumps does not result in
much absorption (unless the accumulated column depth of multiple clumps
produces a significant optical depth in total).  For example, our model
is for $\gamma=1.4$ with $t_2=1000$ and $t_1=0.1$ at $f=1$ (appearing
in the upper panel of Fig.~\ref{fig9}).  At the wavelength of 7~mm,
these upper and lower optical depth limits become approximately 10 and
0.001, respectively.  The mean clump optical depth is thus only $\langle
t \rangle \approx 0.2$, and just 4\% of the clumps have optical depths
in excess of unity along their diameters.  Consequently, the effective
covering factor is closer to $4\% \times 15\% = 0.6\%$.  Even at 15\%,
the shadowing is not severe for W49N-B2; however, this discussion
illustrates how application of this model to other sources may need
to take into account optical effects when interpreting {\it geometric}
covering factors if found to be of order unity or larger.

If the analysis of W49N-B2 is valid, then information about the optical
depth distribution of the clumps has been inferred.  Although optical
depth is the natural variable to be deduced from observables, the
preferred distributions to test against theoretical models are in density
and size.  Of course, the optical depth distribution that is derived
from the data {\it can} be related to the density and size properties of
the clumps.  This is achieved by taking density and size distributions
from theoretical considerations and simulations, and transforming them
to a distribution in $\tau$.  Generally, if the densities and sizes
for the clumps individually obey power laws, then the optical depth
distribution will also be a power law, since $\tau \propto \rho^2\,R$
for uniform clumps.  Consequently, our method that is sensitive to the
optical depth properties of the clumps can help to constrain models that
predict density and size distributions.

\section{DISCUSSION}

The advantage of using uniform, spherical clumps is that the emission
from an individual clump can be derived analytically, and analytic
derivations are useful for facilitating insight into a problem.  On the
other hand, analytic derivations usually arise from assuming some
simplifying conditions.  In this case the sphericity of the clumps is
a simplification, but unlikely to be important enough to change basic
conclusions. Uniformity of the clumps may pose a more severe problem.
In reality, inhomogeneities near hot stars are probably cold, neutral,
and dense cores with a ``skin'' of ionized, photo-evaporating, gas.
It is these ionized layers (plus any diffuse component existing between
the clumps) that produce the observed free-free emission.

At frequencies for which these layers are optically thick, our analysis
based on spherical clumps is equivalent because all that matters is
the projected area of the clumps.  At frequencies for which the layers
are entirely thin, our method predicts the same slope for the flattened
spectrum, but overestimates the scale of the emission (by the factor of
the ratio of the clump volume relative to the layer volume).  But what
is the impact of ionized layers for the intermediate power-law segments
to SEDs that we have derived here?  Qualitatively, the treatment of
uniform spherical clumps allows for a greater range of optical depths
across the face of the clump than for a layer.  That is, if the layers
are geometrically thin relative to the clump size, then the layer is
nearly plane-parallel, so that $\tau(p)$ will be similar across the clump
face.  This suggests that in the case of ionized layers, the turnover in
frequency from the thick $+2$ slope to the thin $-0.11$ slope will occur
over a narrower bandwidth as compared with uniform clumps.  How this
affects the power-law segments is not clear.  The sharper turnover
in the SED of a single clump may lead to better power-law segments
(i.e., segments of more nearly constant slope and less curvature) than
for uniform clumps; however, these segments may exist over a narrower
bandwidth.  This topic needs to be investigated quantitatively.

Also needed is an extension of the results derived here to the case when
the covering factor exceeds unity.  More specifically, the influence of
optical depth effects need to be included for the case of clumps that
``shadow'' other clumps.  As pointed out in the application to the
W49N--B2 source, the covering factor defined in this paper is one of
geometric cross-section only.

It is possible to predict to some degree the influence of optical
depth on the covering factor.  Specifically, $C$ will be a function
of frequency.  At frequencies for which all clumps are thick,
the limit of $C$ defined purely in terms of the accumulated cross-section
of all clumps is valid, and in this limit $C$ achieves its largest value.
However, the SED when all clumps are thick remains a $+2$ power law.
So, when compared to our present treatment that ignores clump
shadowing, the SED shape remains the same for the limit that all
clumps are thick; only the scale of the emission is overestimated,
since the emission from many rearward clumps do not emerge to the
observer.  In other words the greatest level of emission from a region
of size $L$ is $F_\nu = 2\pi\,B_\nu\,L^2/D^2$.

At the opposite extreme, the covering factor drops to zero at frequencies
in which all the clumps are optically thin (or to be more exact,
when the average optical depth through the emitting region of size
$L$ is less than unity).  In this limit geometric shadowing becomes
irrelevant because there is little absorption, and so the results that
we have presented remain valid, {\it both} in the shape of the SED and
the scale.  Combining the thin portion, which is unchanged, with the thick
portion, for which the flux level is overestimated, it appears that the
intermediate power-law segments will prove to be steeper.  Whether or
not these segments will exhibit greater curvature or not is unclear and
must be studied with numerical calculations in the future.

\section{SUMMARY AND MAIN CONCLUSIONS}

We have used expressions for the radio continuum flux density and optical
depth of a single, unresolved, uniform (i.e., temperature and density
are constant), spherical clump to calculate the SEDs for an ensemble
of many such clumps with a power-law distribution of optical depths.
The motivations for this morphology are: (1) the empirical evidence for
clumping over a wide range of scale sizes in ionized and neutral atomic
and molecular phases of interstellar and circumstellar media; and,
(2) the intermediate sloped power-law SEDs observed toward a growing
number of HC HII regions.  The primary thrust of this investigation
was to determine if a power-law SED with slopes intermediate between
the optically thick and thin limits of $\alpha=+2$ and $-0.1$ (where
$F_{\nu}\propto\nu^{\alpha}$) can be understood as a consequence of
emission from a hierarchically clumped medium; and, if so, to investigate
under what conditions intermediate sloped SEDs are formed and over how
large a frequency interval they may occur.

We have found that it is possible for an ensemble of clumps with a
power-law distribution of optical depths to produce power-law SEDs of
intermediate slope over a limited bandwidth.  The frequency interval
over which an intermediate slope holds is controlled by the range of
clump optical depths for an appropriate distribution of clumps N($\tau$).
The greater the range in optical depths, $\tau_{\rm max}$ to $\tau_{\rm
min}$, the broader the bandwidth over which an intermediate slope
persists.  The slope of the intermediate SED power-law is determined
primarily by the parameter $\gamma$ which specifies the fraction of the
nebula that is filled with optically thick clumps at a given frequency.
In our models, intermediate power-law slopes only appear for values of
$\gamma$ between 1 and 2, however, we have not explored all possible
values of parameter space.

We find a good fit to the SED of W49N-B2 using the measured radio
parameters for this source from De Pree \etal\ (2000) for an ensemble of
clumps with $\gamma = 1.5$.  The clump optical depths vary from a maximum
of 300 to 0.3 at 7 mm and the mean optical depth at this wavelength
is 9.5.  The geometric covering factor is $C \approx 0.15$.
The covering factor is small enough to easily produce a low density
halo around the dense ionized core of this HC~HII region, for which
some observational evidence exists.  

Power-law SEDs of intermediate slopes result from the additive
effect of many individual clumps whose turn-over frequencies occur
in sequential order over a limited range in frequency.  The primary
insight gained from our study of hierarchically clumped nebulae is that
the distribution of optical depths of the clump population is solely
responsible for determining the continuum shape, with variations in
size and temperature of the clumps serving only to modulate the level
of the free-free emission.  Logical extensions of this work would be to
investigate the effects introduced by non-spherical clump morphologies,
non-uniform clumps (i.e., variations of temperature and density within
clumps), and clump shadowing.

\begin{figure}
\plotone{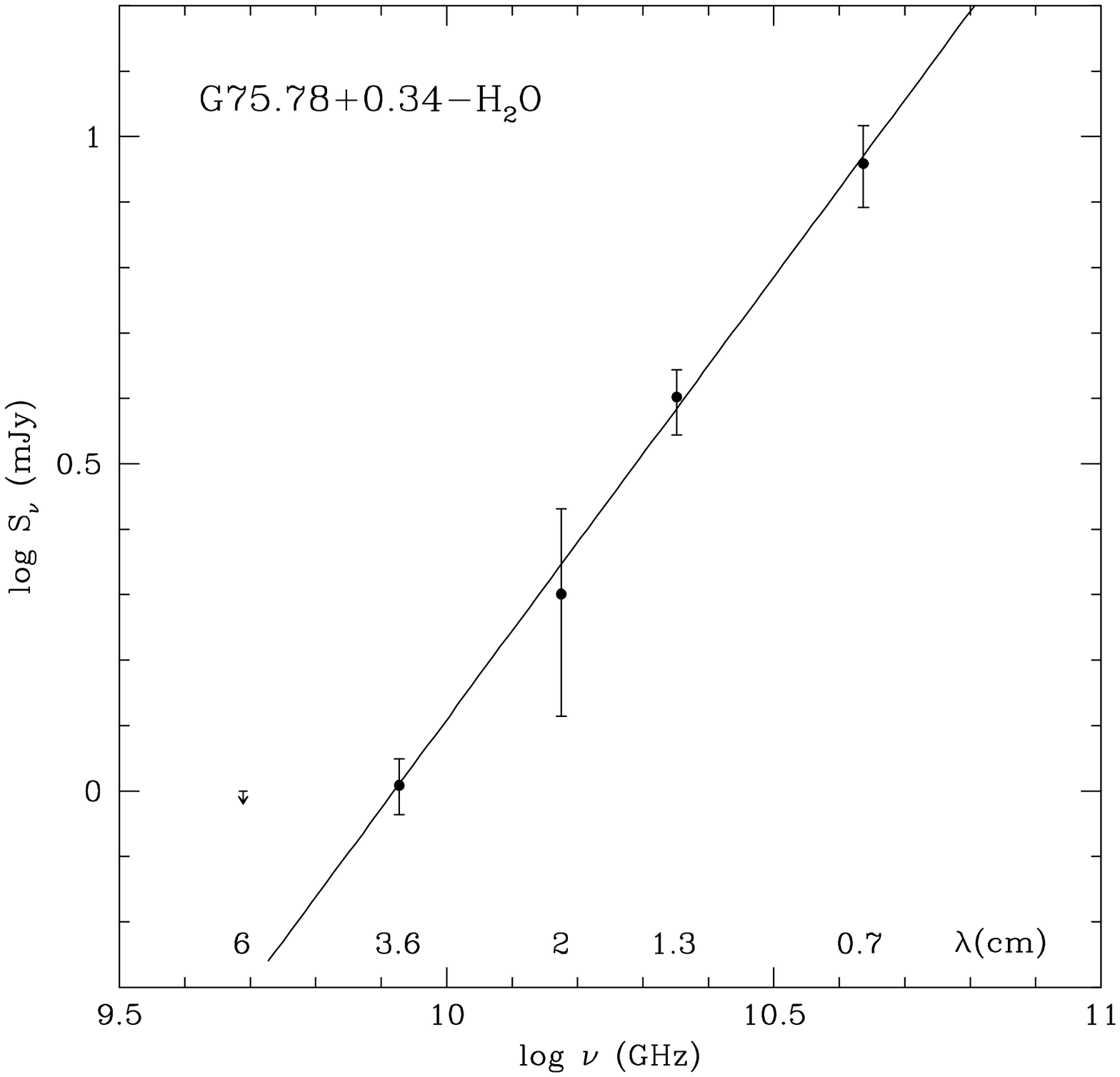}
\caption{
The observed spectral energy distribution for G$75.78+0.34-$H$_2$O.
This object is a good example of an intermediate-sloped power law
between 36 and 7~mm.  It is not detected at 6~cm, likely
because of sensitivity limitations.  This figure is
courtesy of Stan Kurtz (private comm.).
}
\label{fig1}
\end{figure}

\begin{figure}
\plotone{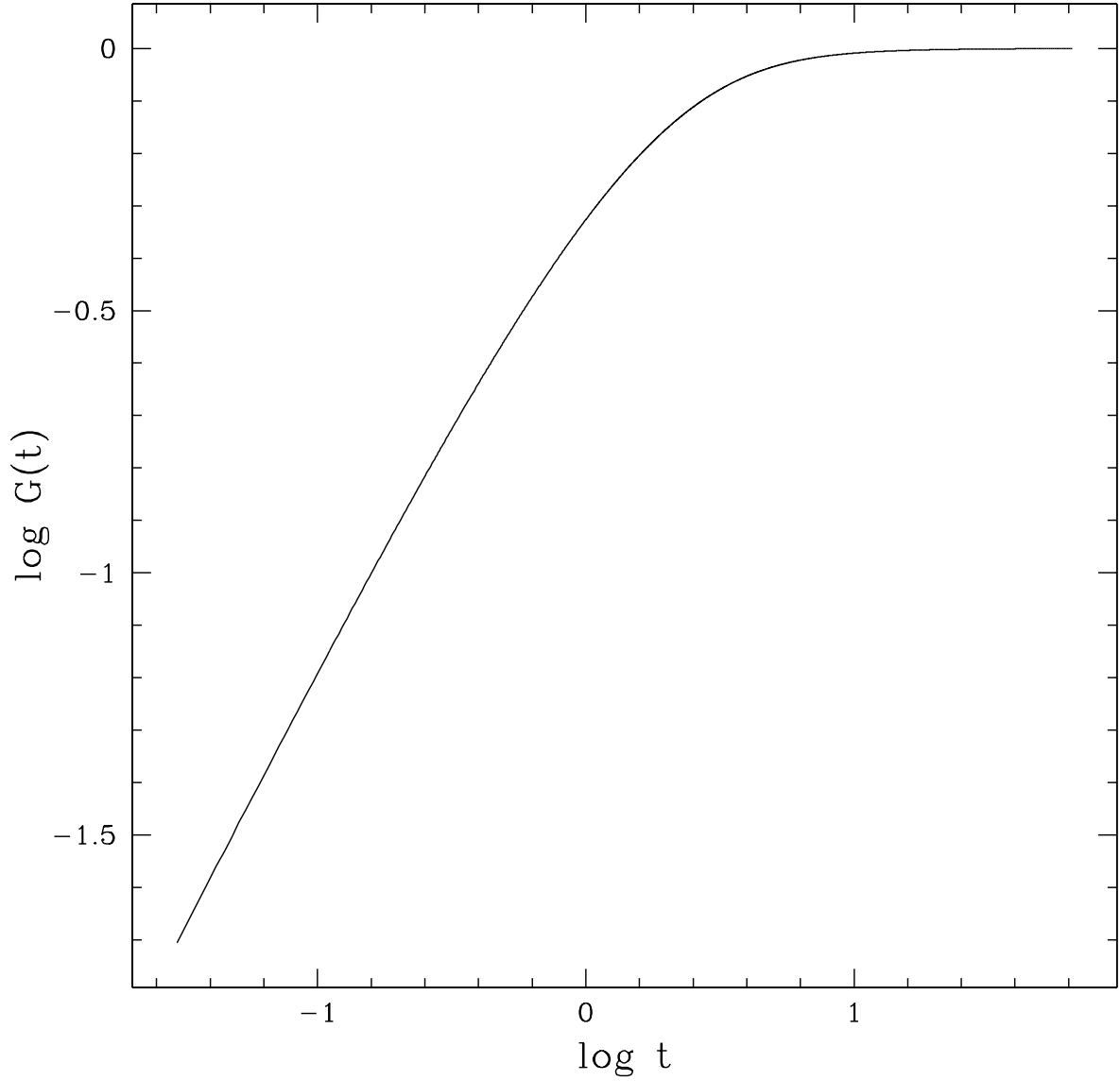}
\caption{A plot of the function $G(t)$.  For low optical
depths, $G \propto t$, but for optically thick clumps,
$G\approx 1$.  The function $G$ represents the relative contribution
to the flux of emission from individual clumps.  Alternatively, it can
be viewed as the spectral contribution of a single clump applying the
transformation $t\propto f^{-2}\,g_\nu$.}

\label{fig2}
\end{figure}

\begin{figure}
\plotone{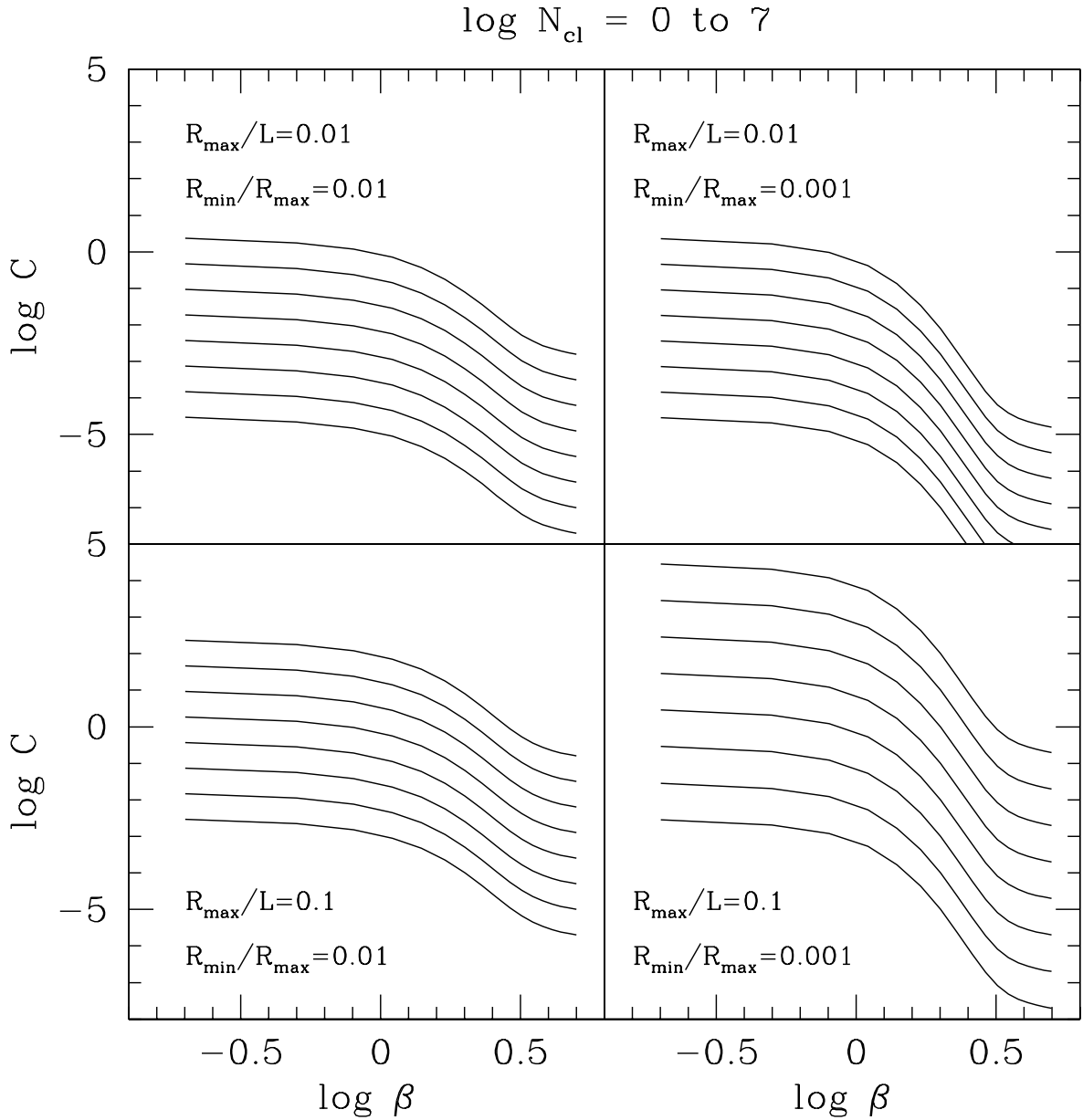}
\caption{A plot of the covering factor $C$ versus the power law
exponent $\beta$ for the distribution of clump sizes.  Each curve is
for a different number of clumps $\log N_{\rm cl}$ from 0 (always the
lowest curve) to 7 (always the highest curve).  The four panels are for
different values of $R_{\rm max}^2 / L^2$ and $R_{\rm min}/R_{\rm
max}$ as labelled.}

\label{fig3}
\end{figure}

\begin{figure}
\plotone{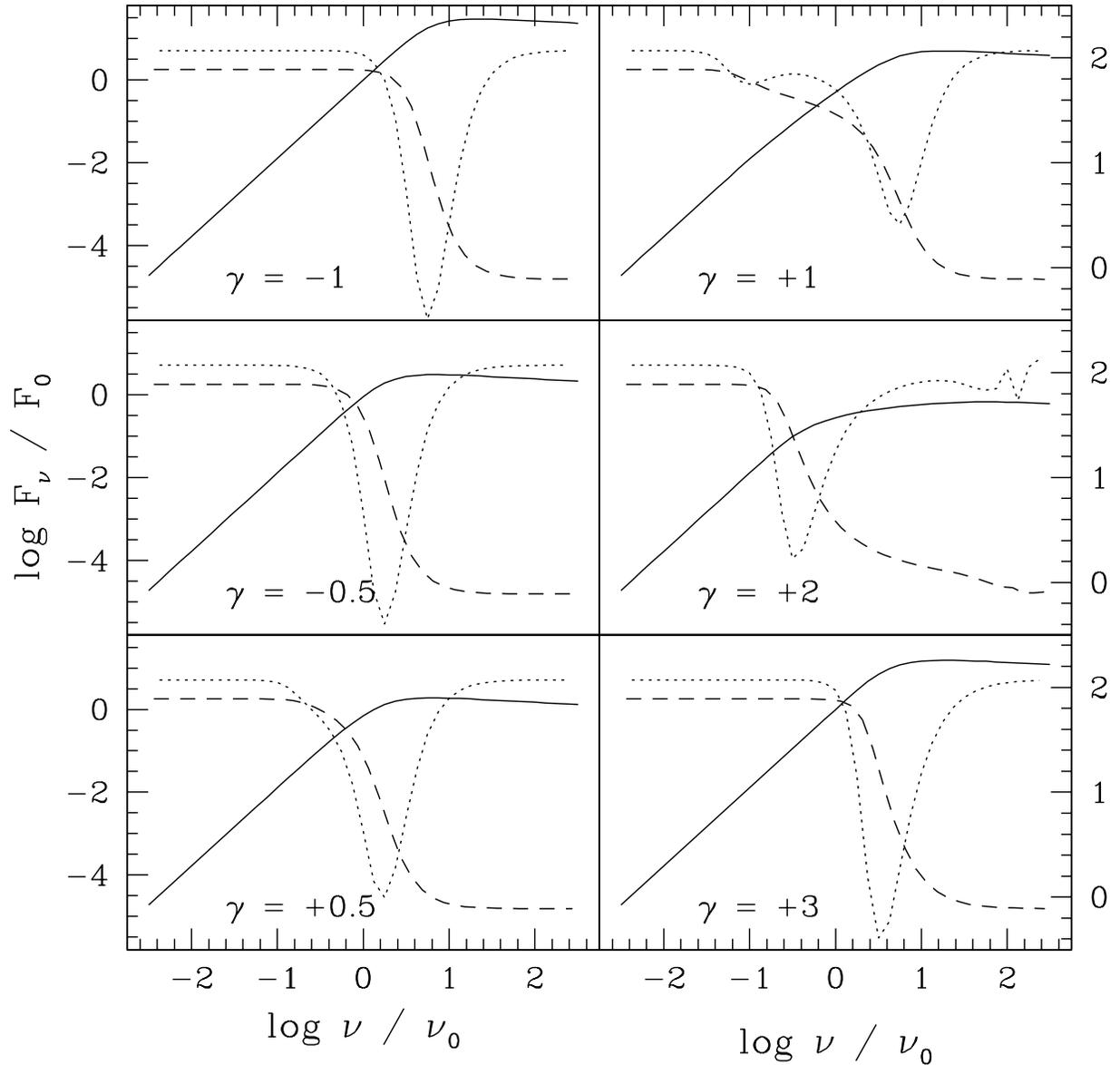}
\caption{A sequence of model SEDs in normalized flux plotted logarithmically
against normalized frequency with $\gamma$ as labelled.  The solid line
is the SED, with axis at far left.  Dashed is the first derivative
of the spectrum, with axis at far right.  Dotted is the second derivative
of the spectrum, which although it has no axis of its own is negative or
near zero because the spectral shapes are concave down.  Downward peaks
in the second derivative curve indicate a transition between power-law
slopes in the SED.  Note that for the middle right panel, the rapid
variations in the first and second derivative curves at far right are
numerical artifacts, and occur where the SED happens to be extremely flat.
(Also seen in Fig.~\ref{fig5}.)  }

\label{fig4}
\end{figure}

\begin{figure}
\plotone{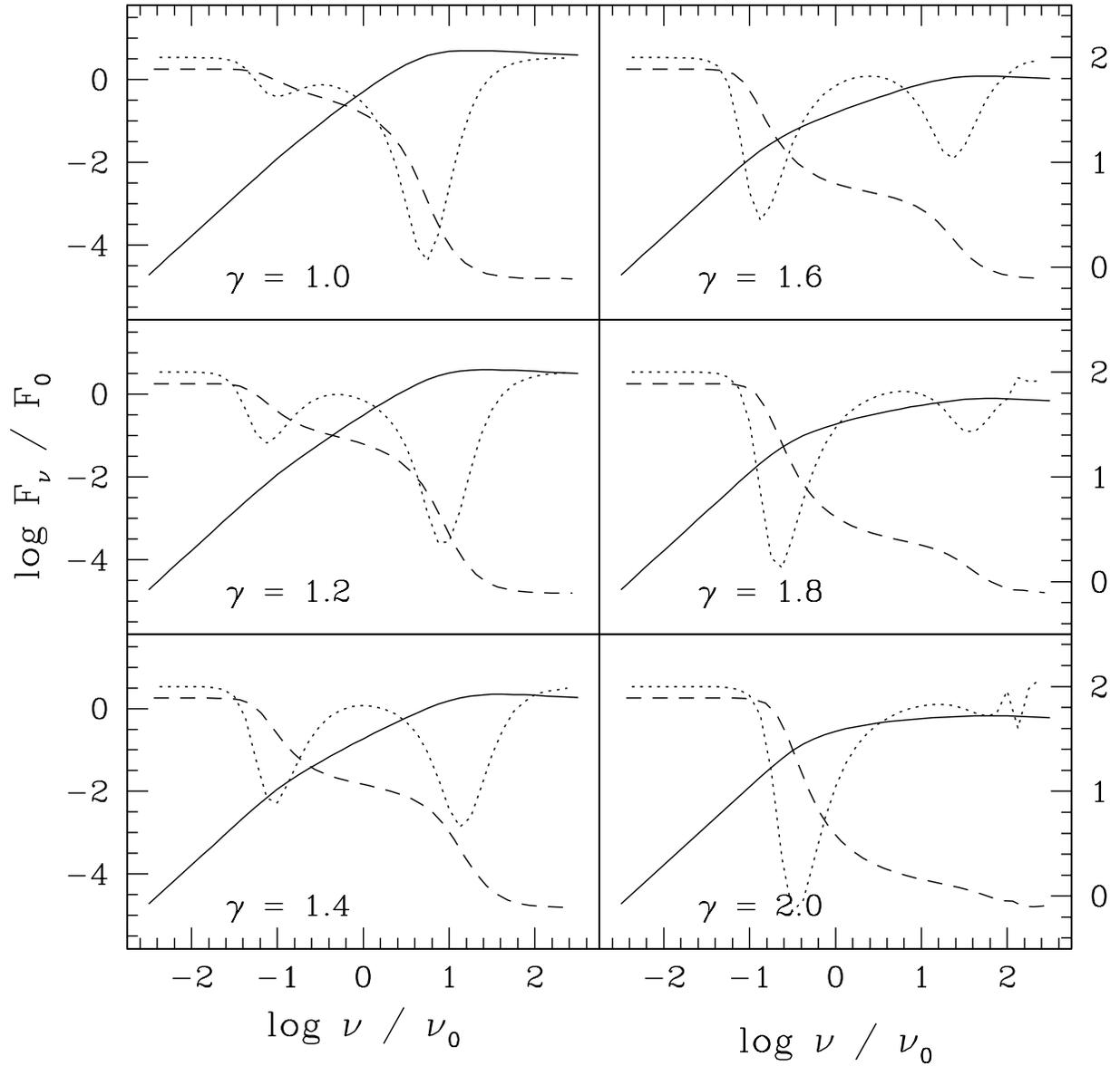}
\caption{A sequence of model SEDs similar to Fig.~\ref{fig4}, now for
a narrower range of $\gamma$ values.  In the range of $\gamma = 1-2$,
a power-law segment with slope intermediate of $+2$ and $-0.11$ exists
over a significant bandwidth.  This ``window'' is demarcated by the two
downward peaks in the second-derivative curve (dotted).  
}

\label{fig5}
\end{figure}

\begin{figure}
\plotone{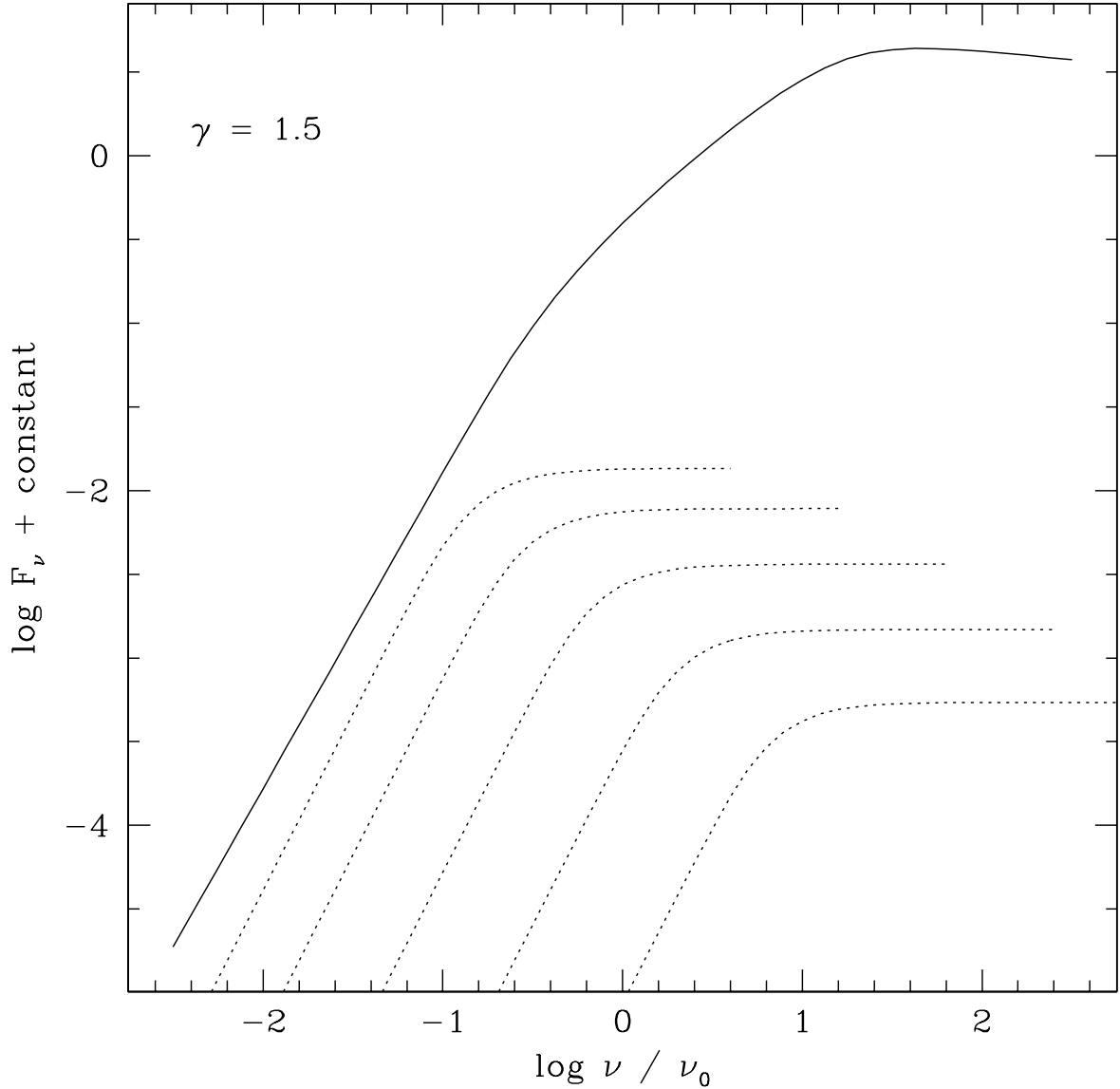}
\caption{An illustration of how intermediate power-laws result.  The solid line
is an SED for $\gamma=1.5$.  The dotted lines are contributions (with an
arbitrary vertical shift) from subgroups of clumps that have different
optical depths at $\nu=\nu_0$.  Although all the dotted curves are similar,
the turnover from a thick spectrum to one that is thin is shifted laterally
from curve to the next, leading in the superposition to an intermediate
power-law slope.  
}

\label{fig6}
\end{figure}

\begin{figure}
\plotone{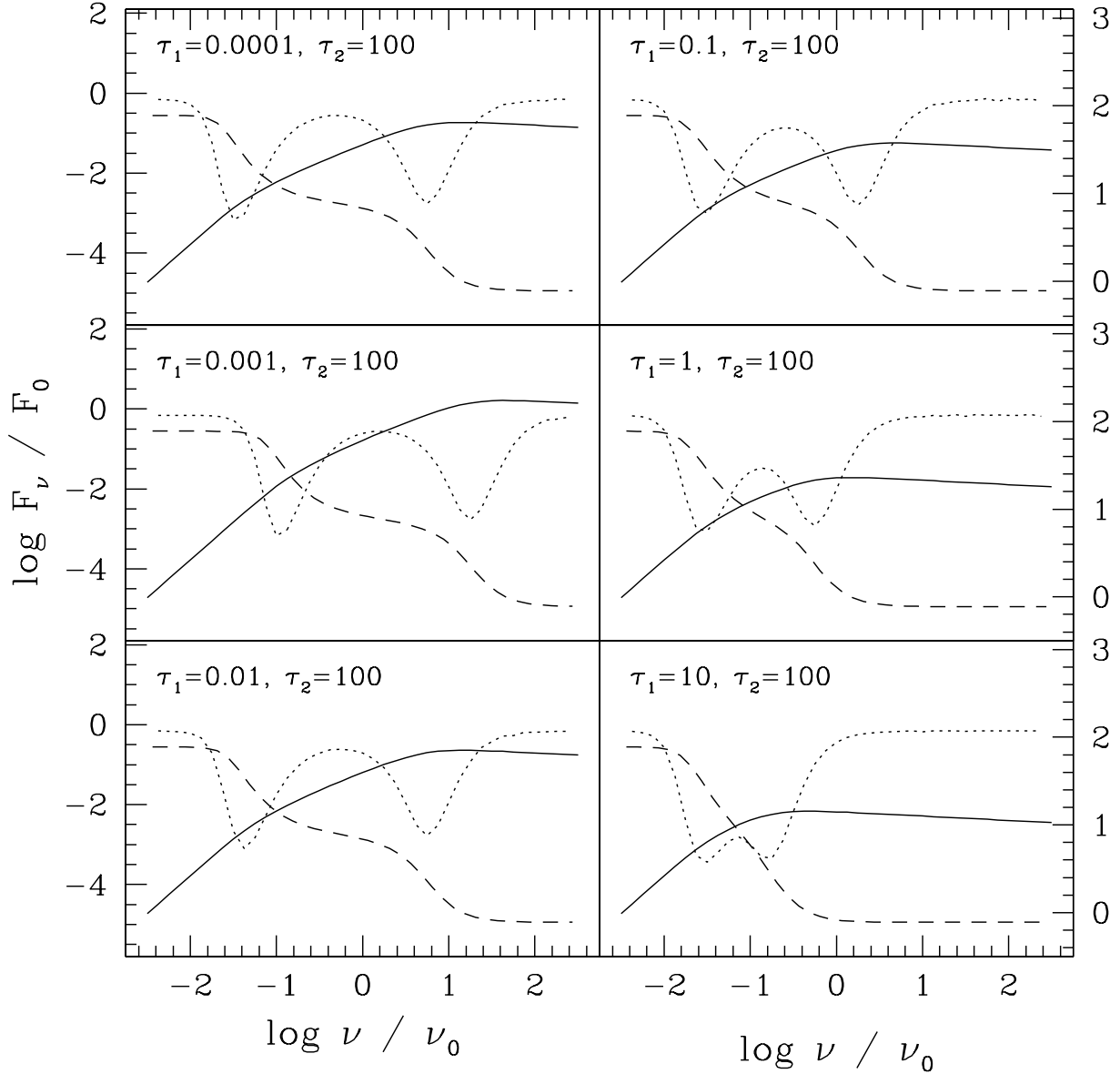}
\caption{Another sequence of model SEDs, similar to Fig.~\ref{fig4},
now with $\gamma=1.5$ fixed.  For the $N(t)$ distribution,
the upper limit $t_2$ is fixed at 100, and the lower limit
$t_1$ is allowed to vary as indicated.  As the lower limit
is increased, the bandwidth for the intermediate power-law segment
narrows and shifts to lower frequency.  The spectrum also drops in
relative brightness.  }

\label{fig7}
\end{figure}

\begin{figure}
\plotone{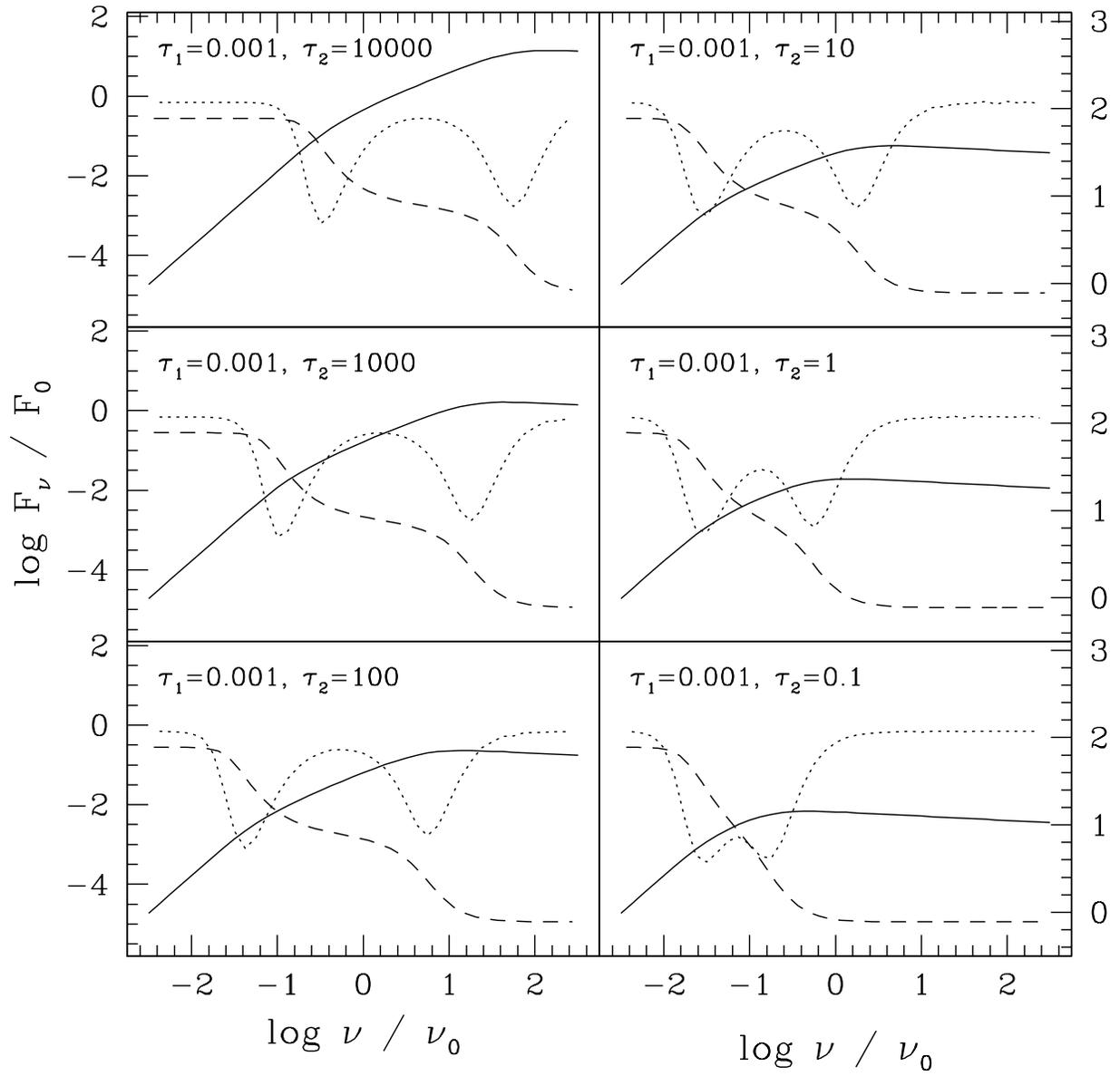}
\caption{Similar to the preceding figure, but now with
$t_1=0.001$ fixed and $t_2$ varied from 10000
down to 0.1 as indicated.  Essentially identical behavior is observed
for the change in the SED properties.}

\label{fig8}
\end{figure}

\begin{figure}
\plotone{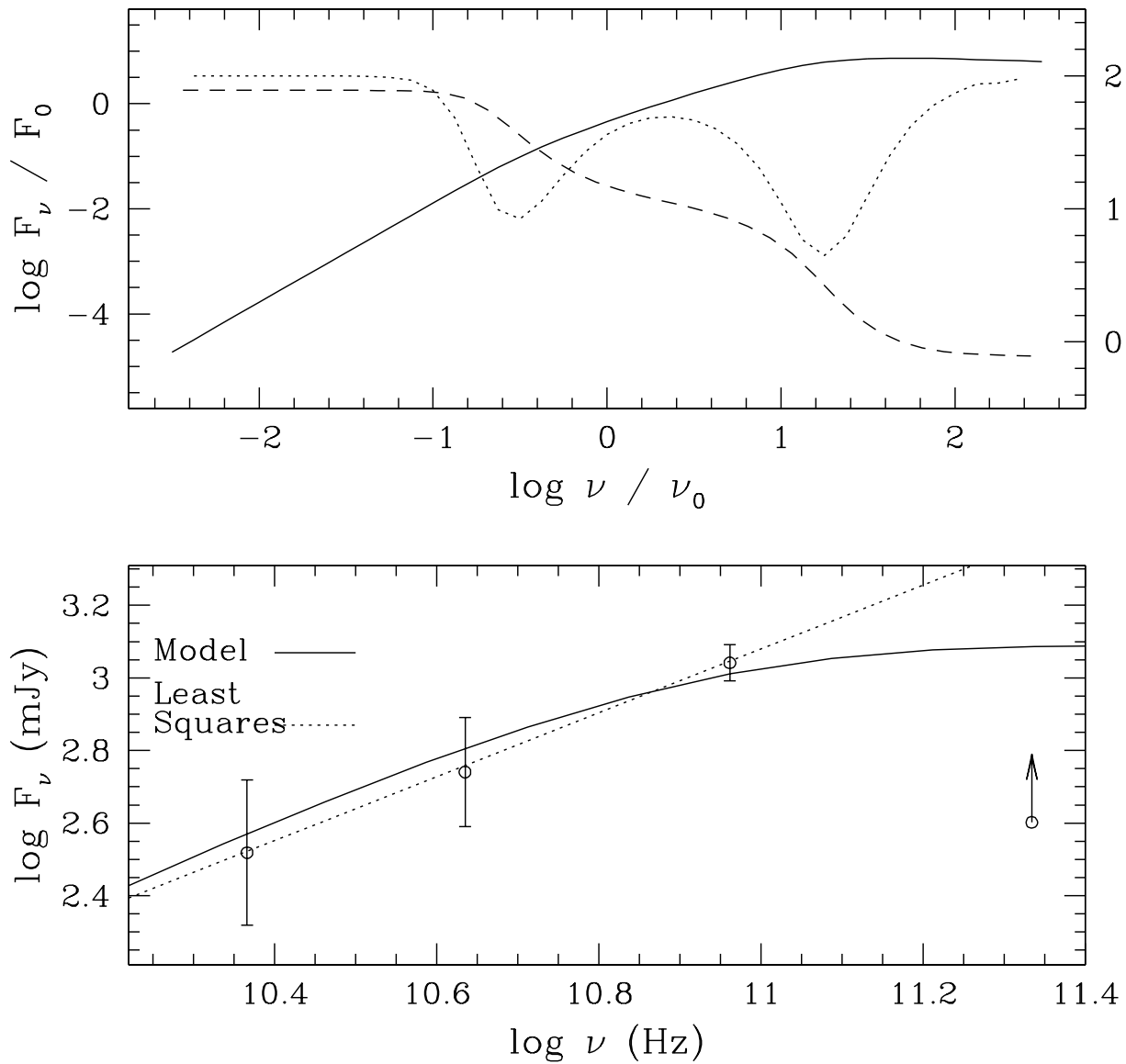}
\caption{Top:  The case of $\gamma=1.4$ yields an intermediate power-law
SED of slope $\approx +1.0$ over a significant frequency bandwidth,
bounded to the right by a turnover to an optically thin spectrum,
and to left by an optically thick one.  Bottom:  Three flux density
points measured for the source W49N--B2, plus a fourth point at right
that is a lower limit (from Wilner \etal\ 2001).  The dotted line
is a power-law fit to the three data points at left and has a slope
of $\alpha=+0.88$.  The solid line is the model calculation from the
upper panel, with $F_0$ and $\nu_0$ chosen to provide for a turnover
to an optically thin spectrum suggested by the lower limit at far right
(also see Tab.~\ref{tab1}).}

\label{fig9}
\end{figure}

\acknowledgements We express thanks to Joe Cassinelli, Ken Gayley,
and John Mathis for helpful comments relating to the emission from
an ensemble of clumps.  We are especially grateful to Stan Kurtz for
providing Figure~\ref{fig1}, and to David Wilner for comments relating
to the 1.4 mm flux density measurement for W49N-B2.  We also express
appreciation to Doug Johnstone and Phil Myers for several valuable
comments on an early draft of this work.  Ignace acknowledges support for
this study from an NSF grant (AST-0241493) and Churchwell acknowledges
partial support for this work from NSF grant AST-9986548.

\begin{table}
\caption[]{Properties of W49N--B2$^a$ \label{tab1}}
\begin{tabular}{ll}
\hline\hline Parameter & Value \\ \hline
RA(J2000) & 19h 10m 13.143s\\
DEC(J2000) & $+09^\circ$ 06' 12.52'' \\
$d$ & 11.4 kpc \\
$F_\nu$(1.4 mm) & $\ge$ 400 mJy \\
$F_\nu$(3.3 mm) & 1100 mJy \\
$F_\nu$(7 mm) & 550 mJy \\
$F_\nu$(13 mm) & 330 mJy \\
SED Slope & $+0.9$ \\ 
$\langle n_{\rm e} \rangle $ & $1.3\times 10^6$ cm$^{-3}$ \\
$EM$ & $2.4\times 10^{10}$ pc cm$^{-6}$ \\
$\dot{N}_\gamma$ & $2\times 10^{49}$ photons s$^{-1}$ \\
$L$ & 0.007 pc \\ \hline 
\end{tabular}

\vspace{1ex}

$^a$ {\small See De~Pree \etal\ (2000) for errors on all quantities. }

\end{table}

\end{document}